**Review Article**

# Thin film growth of MAX phases as functional materials

Abhijit Biswas[1],* Varun Natu[2], and Anand B. Puthirath[1]

[1]Department of Materials Science and Nanoengineering, Rice University, Houston, TX 77005, USA

[2]Department of Materials Science & Engineering, Drexel University, Philadelphia, PA 19104, USA

## ABSTRACT

Layered nanolaminate ternary carbides, nitrides and carbonitrides with general formula $M_{n+1}AX_n$ or MAX ($n$ = 1, 2, or 3, M is an early transition metal, A is mostly group 13 or 14 element, and X is C and/or N) has revolutionized the world of nanomaterials, due to the coexistence of both ceramic and metallic nature, giving rise to exceptional mechanical, thermal, electrical, chemical properties and wide range of applications. Although several solid-state bulk synthesis methods have been developed to produce a variety of MAX phases, however, for certain applications, the growth of MAX phases, especially in its high-quality epitaxial thin films form is of increasing interest. Here, we summarize the progress made thus far in epitaxial growth and property evaluation of MAX phase thin films grown by various deposition techniques. We also address the important future research directions to be made in terms of thin-film growth. Overall, in the future, high-quality single-phase epitaxial thin film growth and engineering of chemically diverse MAX phases may open up interesting new avenues for next-generation technology.

Correspondence e-mail: **01abhijit@gmail.com**, **abhijit.biswas@rice.edu**

**(ORCID:** 0000-0002-3729-4802**)**

## INTRODUCTION

Layered nanolaminate ternary carbides and nitrides with chemical formula $M_{n+1}AX_n$ or MAX ($n$ = 1, 2, or 3, M is an early transition metal, A is a group 13 or 14 element, and X is C and/or N) is a unique class of material showing a combination of both ceramic and metallic properties [1-3]. Because of this unique nature, nanolaminate MAX materials show high electrical and thermal conductivity, excellent mechanical properties e.g. high Young's modulus and Vickers hardness, low frictional coefficient, high oxidation resistance, high damage tolerance than other ceramics, high thermal shock resistance, corrosion resistance, electro-catalysis for renewable energy production and high radiation damage tolerance for space and nuclear applications [4, 5]. Generally, MAX forms a layered hexagonal crystal structure (space group $P6_3/mmc$), where layers of edge-shared $M_6X$-octahedra separated by a two-dimensional (2D) closed packed layers of "A" elements, which are located at the center of trigonal prisms forming an alternate stacking of M-X and A layers. Its high metallicity (low resistivity ~μΩ-cm) is due to the presence of a substantial amount of density of states at the Fermi level ($E_F$), with a dominant contribution coming from the *d-d* orbitals of the M-element. The M-X bonds are extremely strong because of the presence of both covalent and metallic bonding nature, where M-A bonds are relatively weak, responsible for layered structure as well as a unique combination of both ceramic and metallic characteristics, thus bridging the gap between metals and ceramics. The MAX phase can be categorized further into 211 ($M_2AX$), 312 ($M_3AX_2$) and 413 ($M_4AX_3$) phases (**Fig. 1**). In 1963, the first discovery of the MAX phase was reported by Nowotny's group [6]. However, MAX remained largely unexplored for almost three decades, until the successful synthesis of bulk phase $Ti_3SiC_2$ MAX by Barsoum and El-Raghy [7], and henceforth ~160 new compounds have been synthesized in its bulk form, revolutionizing the world of layered functional carbides and nitrides nanomaterials [8].

One of the remarkable features of MAX is the conversion of it to MXene, formation of a new layered 2D material, analogue to graphene, showing a wide-range of functional properties and enormous potentials in almost every application sector [9]. 2D MXenes are typically obtained from 3D solid MAX via selective etching of A-layers by using fluoride ion containing acidic solutions. However, the surfaces of these layered 2D sheets are chemically terminated/functionalized with a different/selective group ($M_{n+1}X_nT_x$, where $T_x$ is the surface termination) originating from the etchant, affecting the material properties extensively, e.g. electrical conductivity [10], and thus hindering their usefulness for clean device fabrications for high-performance nano-electronics, embodying clean non-chemical fabrication protocols.

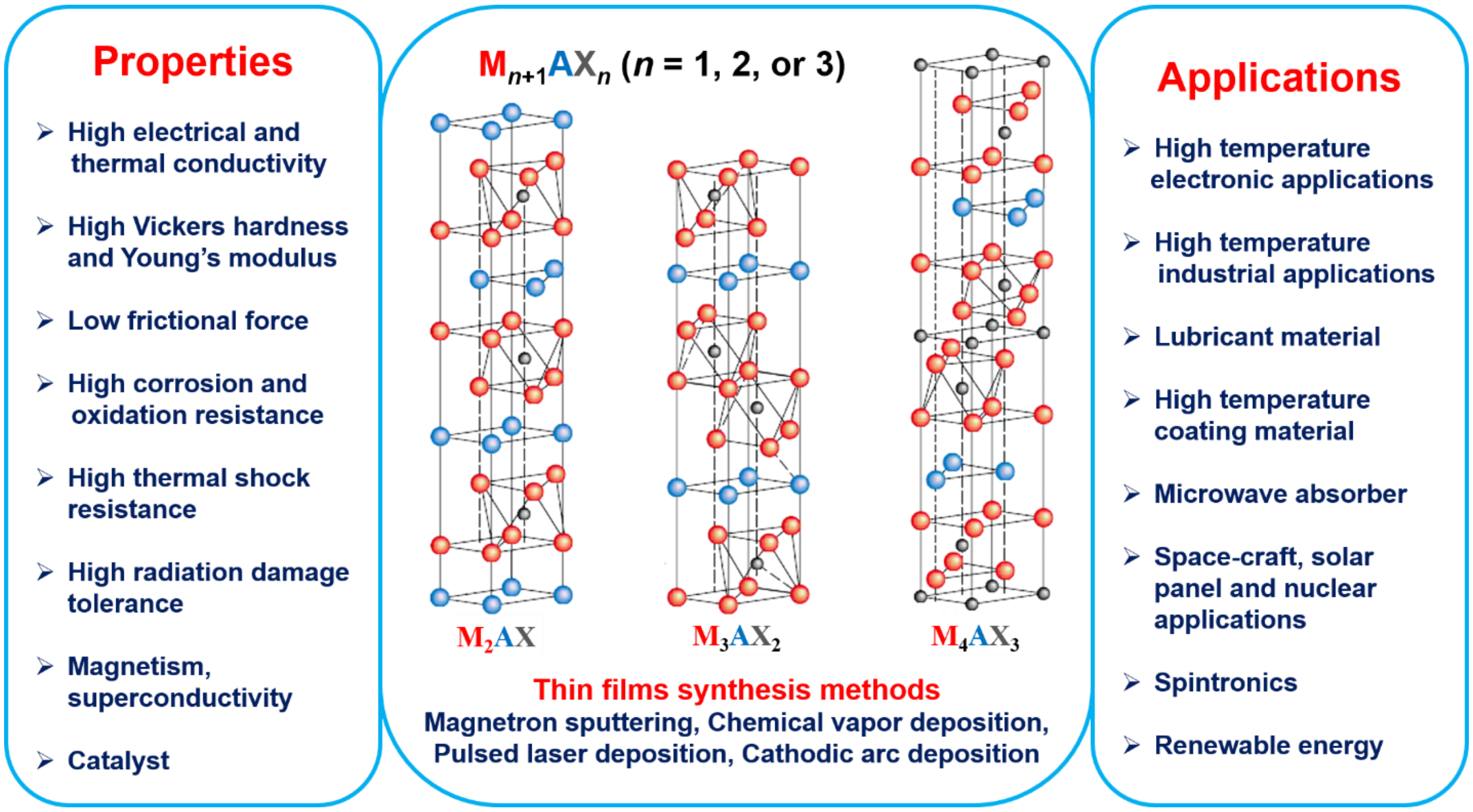

**Figure 1:** Layered ternary $M_{n+1}AX_n$ ($n$ = 1-3) with a wide-range of properties and applications. The crystal structures were adapted with permission from Barsoum and El-Raghy, *American Scientist* 2001; **89**: 334, Copyright 2001 Sigma Xi, The Scientific Research Honor Society [2].

Although, a large numbers of layered bulk MAX phase compounds have been discovered (**Table 1**), as said however for relevant technological importance, the growth of MAX phases in its high-quality single-phase epitaxial thin film form of these films are highly desired. Therefore, in the last couple of decades, numerous attempts have been made to grow epitaxial MAX phase thin films by using various thin film growth methods and there are some review articles published along these directions [11-13]. Therefore, considering the growing interest about MAX phase thin films, in this article we attempt to summarize the progress made thus far in the growth and property evaluation of most, if not all the MAX phase thin films ($n$ =1-3). It was seen that epitaxial thin films also show excellent electrical, thermal, mechanical, tribological, magnetic, oxidation resistance, and corrosion properties with usefulness in high temperature based industrials applications, lubricant and coating materials, as well as radiation damage tolerance nuclear materials, solar panels, and space-craft, making MAX based devices huge potential application worthiness. We also address several challenges and future explorations to be made in terms of thin film growth. Hence, growth and engineering of MAX phases especially in its high-quality epitaxial thin film form is important, showing unique functionalities and occasionally comparable or even suppressing the MXene properties, making MAX as next-generation application-worthy material.

**Table 1:** List of synthesized bulk $M_{n+1}AX_n$ ($n$ = 1-3) compounds. This is an updated version of the earlier list provided by Sokol *et al.*, *Trends in Chemistry* 2019; **1**: 210 [8].

| Phase | Materials |
|---|---|
| **211** ($M_2AX$) | $Ti_2AlC$, $V_2AlC$, $Cr_2AlC$, $Nb_2AlC$, $Ta_2AlC$, $Zr_2AlC$, $Hf_2AlC$, $Ti_2GaC$, $V_2GaC$, $Cr_2GaC$, $Nb_2GaC$, $Mo_2GaC$, $Ta_2GaC$, $Mo_2AlC$, $Mo_2GaC$, $Mn_2GaC$, $Sc_2InC$, $Ti_2InC$, $Zr_2InC$, $Nb_2InC$, $Hf_2InC$, $Ti_2SnC$, $Zr_2SnC$, $V_2SnC$, $Nb_2SnC$, $Mo_2AuC$, $Nb_2CuC$, $Ti_2CdC$, $Sc_2InC$ $Ti_2InC$, $Zr_2InC$, $Nb_2InC$, $Hf_2InC$, $Ti_2TlC$, $Zr_2TlC$, $Hf_2TlC$, $Ti_2PbC$, $Zr_2PbC$, Hf2PbC, $Ti_2GeC$, $V_2GeC$, $Cr_2GeC$, $Nb_2GeC$, $V_2PC$, $Nb_2PC$, $Ti_2SC$, $Zr_2SC$, $Nb_2SC_{0.4}$, $Nb_2SC$, $Hf_2SC$, $Ti_2ZnC$, $V_2ZnC$, $V_2AsC$, $Nb_2AsC$, $Ti_2GaN$, $Cr_2GaN$, $V_2GaN$, $Mo_2GaN$, $Ti_2InN$, $Zr_2InN$, $Hf_2SnN$, $Ti_2InN$, $Zr_2InN$, $Zr_2TlN$, $Ti_2AlN$, $Zr_2SeC$, $Ti_2AuN$, $Ti_2ZnN$, $(Cr,Mn)_2GaC$, $(Cr,Mn)_2GeC$ $(Ti_{0.5}Nb_{0.5})_2AlC$, $(Ti_{0.5}V_{0.5})_2AlC$, $(Mo_{4/3}Y_{2/3})AlC$, $(Nb_{0.5}V_{0.5})_2AlC$, $(W_{2/3}Sc_{1/3})_2AlC$, $(W_{2/3}Y_{1/3})_2AlC$, $(Mo_{2/3}RE_{1/3})_2AlC$ (RE= Nd, Gd, Tb, Dy, Ho, and Er), $Ti_2(Al_xCu_{1-x})N$, $V_2(A_xSn_{1-x})C$ (A = Fe, Co, Ni, Mn), $(Ti_{1/5}V_{1/5}Zr_{1/5}Nb_{1/5}Ta_{1/5})_2AlC$ |
| **312** ($M_3AX_2$) | $Ti_3AlC_2$, $Ta_3AlC_2$, $Zr_3AlC_2$, $Ti_3SiC_2$, $Ti_3GaC_2$, $Ti_3GeC_2$, $Ti_3InC_2$, $Ti_3SnC_2$, $Zr_3SnC_2$, $Hf_3SnC_2$, $Ti_3AuC_2$, $Ti_3IrC_2$, $Ti_3ZnC_2$, $Ti_3ZnC_2$, $(Cr_2V)AlC_2$, $(Cr_2Ti)AlC_2$, $(Ti_2Zr)AlC_2$, $(Mo_2Sc)AlC_2$, $(Mo_2Ti)AlC_2$, $(V,Mn)_3GaC_2$ $(V_{0.5}Cr_{0.5})_3AlC_2$, $(Zr_{0.5}Ti_{0.5})_3AlC_2$, $(Ti_{0.5}V_{0.5})_3AlC_2$, $(Cr_{2/3}Ti_{1/3})_3AlC_2$, $(Ta_{1-x}Ti_x)_3AlC_2$, $(Cr_{1/3}Ti_{2/3})_3AlC_2$, $Ti_3Al(C_{0.5}N_{0.5})_2$, $Ti_3(Al_xCu_{1-x})C_2$ |
| **413** ($M_4AX_3$) | $Ti_4SiC_3$, $Nb_4SiC_3$, $Ti_4AlN_3$, $Ta_4AlC_3$, $Nb_4AlC_3$, $V_4AlC_3$, $Ta_4GaC_3$, $Ti_4GaC_3$, $(Mo_2Ti_2)AlC_3$, $(Cr_2V_2)AlC_3$, $(V_{0.5}Cr_{0.5})_4AlC_3$, $(Nb_{0.5}V_{0.5})_4AlC_3$, $(Cr_{5/8}Ti_{3/8})_4AlC_3$ $(TiVNbMo)AlC_3$, $(TiVCrMo)AlC_3$, |

## THIN FILM GROWTH METHODS

Here we briefly summarize the various growth techniques that have been used for the deposition of MAX phases thin films. These are mainly chemical vapor deposition (CVD), magnetron sputtering, cathodic arc deposition, and pulsed laser deposition (PLD) [11]. In principle, each of these techniques has its own advantages and disadvantages. Generally, CVD requires very high temperature (>1000 °C), however giving little success as several solution-based precursors are generally used, and leaving lots of residues on the surface. High power dc or radio frequency (rf) magnetron sputtering has been most often used to grow MAX phase films. Its advantage is that individual target sources are used to sputter, giving control over each element and the target compositions. In this technique films are grown typically at temperature ~700-1000 °C [11]. In some cases, reactive magnetron sputtering has also been employed to deposit the films by using $N_2$ as a source. The cathodic arc deposition technique has also been used to grow some MAX phase thin films. This method allows to grow films at relatively low substrate temperature of ~500 °C, as a high degree of ionization of all the species (deposition flux) provides the control of ion energy, allowing the films to grow at relatively lower temperatures [11]. By using the advantage of the high-energy vapor phase deposition, PLD technique has also been employed to grow various MAX phase thin films. Here one can use a single stoichiometric bulk target of the desired MAX phase for ablation by high energy pulsed laser beam (~eV), generating the plasma and thereby the true stoichiometric transfer of all the species on the substrate surface and thus providing a high possibility in the formation of a single phase epitaxial films. By PLD, films are typically grown at ~700 °C. Starting with the first attempt of Ti-Si-C thin film growth by PLD by Phani *et al.*, [14], till now several attempts have been made to grow MAX phase thin films as summarized later on, however further investigation is still required to achieve the highest-crystalline quality of films.

## MAX PHASE THIN FILMS

### 1. $M_2AX$ (*n*=1)

211 MAX phases are the first layered compound of the ternary MAX series, showing excellent metallic behavior and high sustainability in extreme environments (e.g. high temperature, corrosive, oxidizing). Their large *c*/*a* lattice constant ratio, makes the properties of the material highly anisotropic [8]. A typical band structure calculation shows the reasonable amount of density of states (DOS) at the Fermi level ($E_F$) [15], with a majority of the charge carriers (*p*-type) coming from the Ti-3*d* orbitals (**Fig. 2a**). Therefore, growing thin films of 211 MAX phases became an interesting avenue among the materials synthesis community. Wilhelmsson *et al.*, for the first time reported the epitaxial growth of $Ti_2AlC$ MAX phase thin film by using the magnetron sputtering deposition technique [16]. They grew films at ~900 °C on $Al_2O_3$ (000l) substrates with a ~20 nm TiC (111) seed buffer layer for improving the epitaxy, and measured the resistivity of ~44 μΩ-cm at room temperature. Later they also measured the hardness and Young's modulus of ~600 nm $Ti_2AlC$ film and obtained the values of ~20 GPa, and ~260 GPa, respectively [17]. High current pulsed cathodic arc deposition method was employed by Rosén *et al.*, to grow epitaxial $Ti_2AlC$ thin films on sapphire and they observed a (001) ($Ti_2AIC$) // (0001) ($Al_2O_3$), and (11−20) ($Ti_2AlC$) // (11−20) ($Al_2O_3$) epitaxial relationship by using high-resolution HRTEM [18]. Pshyk *et al.*, grew ~120 nm $Ti_2AlC$ MAX phase thin films by using electron beam physical vapor deposition at 700 °C [19] and investigated structural, nano-mechanical and tribological properties and found the hardness and elastic modulus of ~4.8 GPa and ~182.5 GPa, respectively (**Fig. 2b**). The elastic modulus is found to be lower than in case of films grown by sputtering [17], possibly due to used low growth temperature (~700 °C), leading to the formation of solid solution of Al in TiC. Films also showed a very low coefficient of friction of ~0.3, with applied load of ~500 μN (**Fig. 2c**). Frodelius *et al.*, tried to optimize the growth of the $Ti_2AlC$ phase by sputtering and found that additional Ti flux is important to grow the stoichiometric 211 MAX phase [20]. Later they also studied the oxidation of $Ti_2AlC$ film grown at ~500 °C, forming a $TiO_xC_y$ layer at the surface [21]. This shows that high temperature is important to grow pure phase MAX films and also to obtain a good oxidation resistance. Electrical resistivity measurement by Mauchamp *et al.*, shows the anisotropic behavior with room temperature resistivity of ~36 μΩ-cm for $Ti_2AlC$ thin films [15]. $Ti_2AlC$ MAX phase was also deposited on cubic MgO (001) substrates by magnetron sputtering. However, the films were found to be polycrystalline with hexagonal laminar grain with lateral size between ~150-400 nm [22]. Feng *et al.*, also found that high temperature annealing is important in the formation of high-quality $Ti_2AlC$ phase thin films [23].

Regarding other 211 MAX phases, Högberg *et al.*, reported the growth of epitaxial $Ti_2GeC$ thin films on $Al_2O_3$ (0001) substrates at ~1000 °C by using the magnetron sputtering technique [24]. After that they also grew $Ti_2AlN$ thin films and observed low electrical resistivity of ~39 μΩ-cm, comparable to the bulk value [25]. Emmerlich *et al.*, reported the growth of $Ti_2GeC$ and $Ti_2SnC$ thin films showing low electrical resistivity of ~15-50 μΩ-cm [26]. Moreover, $Ti_2GeC$ films show high carrier density of ~$3.5\times10^{27}$ $m^{-3}$ [27]. Beckers *et al.*, grew $Ti_2AlN$ films on MgO (111) and $Al_2O_3$ (0001) substrates by reactive sputtering methods and found that film decomposes at ~800 °C, because of the onward diffusion of Al [28]. Nanolaminate $Ti_2AlN$ thin films were also synthesized by the multilayer deposition of ~4.5 nm Ti layers and ~3 nm thin AlN layers on $Si/Si_3N_4$ substrate and subsequent rapid thermal annealing [29]. A combined cathodic arc/sputter technique was also employed to grow dense and high-temperature stable $Ti_2AlN$ MAX films [30].

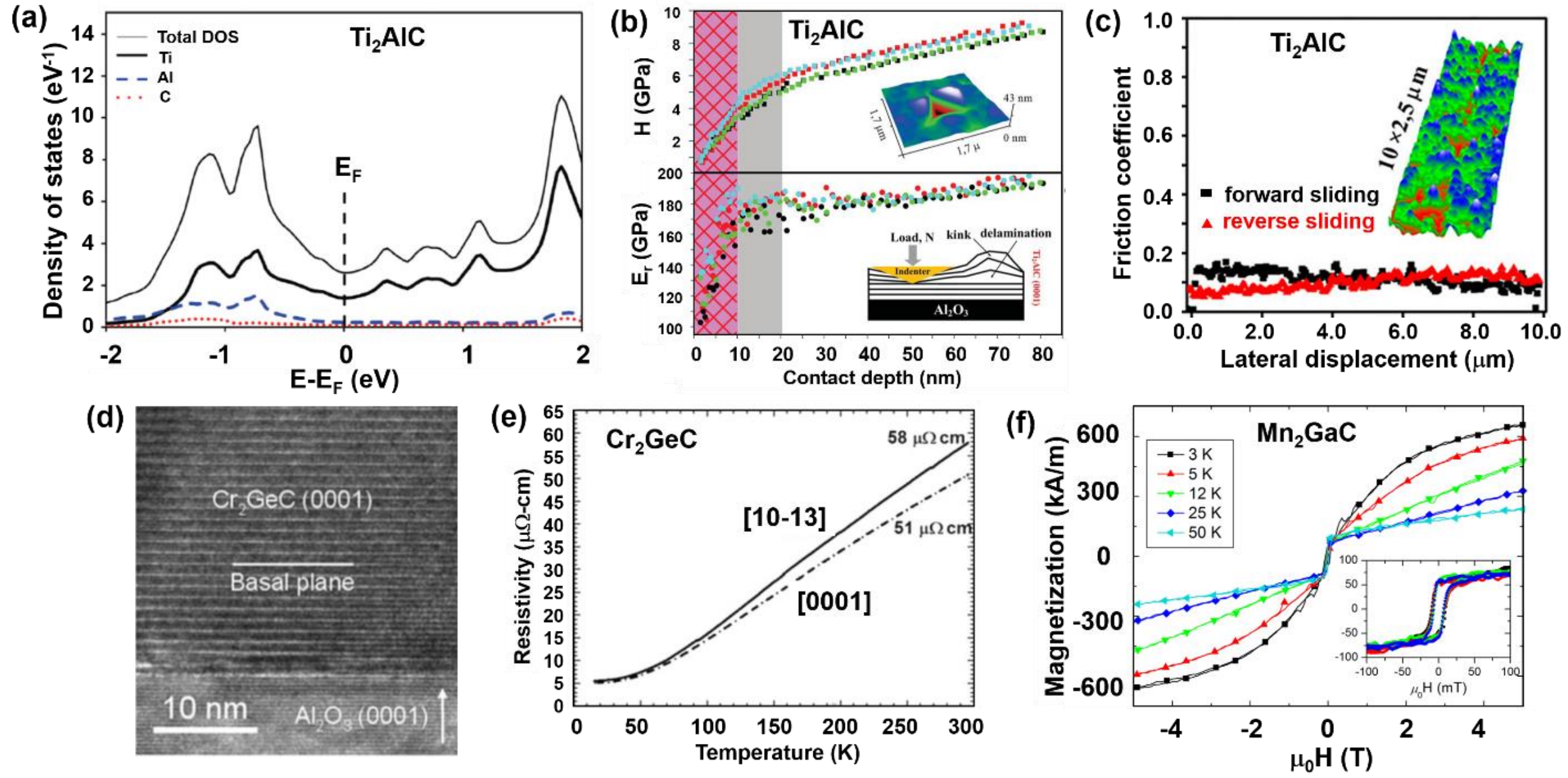

**Figure 2: Properties of $M_2AX$ (*n*=1) phases:** (a) Theoretical band structure calculation of $Ti_2AlC$ shows the presence of non-zero density of states (DOS) at the Fermi level ($E_F$), with dominant contribution from Ti-3*d* orbitals. (b), (c) Hardness and elastic moduli vs. contact depth of a ~120 nm thick $Ti_2AlC$ film grown on sapphire. Inset shows the scanning probe images after the nano-indentation depth of 10 mN. (d) High-resolution TEM image shows the epitaxial growth of $Cr_2GeC$ thin films along the basal plane. (e) Resistivity measurement of a ~190 nm $Cr_2GeC$ thin film on $Al_2O_3$ (0001) shows anisotropic metallic behavior. (f) In-plane magnetization measurement of ~100 nm $Mn_2GaC$ thin film grown on MgO (111) shows the presence of ferromagnetic ordering below 230 K. Inset shows the clear magnetic hysteresis loop at different temperatures. Adapted

with permission from Mauchamp *et al.*, *Phys Rev B* 2013; **87**: 235105, copyright 2013 American Physical Society [15]; Pshyk *et al.*, *Mater Res Lett* 2019; **7**: 244–250, copyright 2019 The Author(s). Published by Informa UK Limited, trading as Taylor & Francis Group [19]; Eklund *et al.*, *Phys Rev B* 2011; **84**: 075424, copyright 2011 American Physical Society [32]; Dahlqvist *et al.*, *Phys Rev B* 2016; **93**: 014410, copyright 2016 American Physical Society [40].

Interestingly, Scabarozi *et al.*, grew $Nb_2AlC$ thin film on sapphire by magnetron sputtering and remarkably, it shows the superconducting transition at ~440 mK [31]. This is a unique example of MAX phase thin films which shows superconductivity. Eklund *et al.*, investigated the epitaxial growth and electrical transport of ~190 nm $Cr_2GeC$ thin film by growing them on $Al_2O_3$ (0001) substrate by sputtering [32]. Before the growth, a seed layer of TiN (111) was pre-deposited on sapphire. Pure-phase films were found to be epitaxial as confirmed by the HRTEM (**Fig. 2d**). Moreover, electrical transport shows anisotropic behavior with resistivity of ~51 μΩ-cm (along 0001) and ~58 μΩ-cm (along 10-13) (**Fig. 2e**). Later, Scabarozi *et al.*, doped V into the film and grew $(Cr_{1-x}V_x)_2GeC$ and investigated the electronic and elastic properties [33]. It shows metal-like electrical conductivity with a resistivity of ~50 μΩ-cm, as well as a minimal effect on the elastic moduli even after alloying.

$Cr_2AlC$ films were synthesized by depositing multilayers of Cr, C, and Al on $Si/Si_3N_4$ substrates showing reasonable mechanical properties [34]. Hopfeld *et al.*, measured the tribological properties of $Cr_2AlC$ films, obtaining the friction coefficients ranging between ~0.30-0.70 [35]. Recently, epitaxial thin films of $Cr_2AlC$ on MgO (111) and $Al_2O_3$ (0001) substrates by pulsed laser deposition show the thickness dependent semiconductor-like and metal-like behavior, suggesting a percolation thickness for the observation of electronic properties [36]. Very recently, dual phase $Cr_2AlC$ thin films were grown on Si (100) substrate by magnetron sputtering showing excellent radiation-tolerance behavior, useful for future fusion and fission nuclear reactors [37]. Stelzer *et al.*, measured the high-temperature resistivity of polycrystalline $Cr_2AlC$ thin films and observed the changes in resistivity due to the structural changes from amorphous to a hexagonal disordered solid solution structure and from the latter to MAX phase [38]. Jiang *et al.*, gave an effort to grow $V_2AlC$ thin films by magnetron sputtering [39]. Dahlqvist *et al.*, grew $Mn_2GaC$ thin films on MgO (111) substrates and investigated their magnetic properties, showing magnetically driven anisotropic structural changes and ferromagnetic ordering below ~230 K (**Fig. 2f**), with robust intra-layer ferromagnetic spin coupling [40]. Considering the antiferromagnetic ground state of $Cr_2GaC$, Petruhins *et al.*, successfully synthesized thin films of this compound [41].

From another application point of view, $Ti_2AuN$ thin films were obtained from Au-covered $Ti_2AlN$ thin films grown via magnetron sputtering on $Al_2O_3$ (000*l*) substrate and subsequent high-temperature annealing which induces the intercalation of noble metals into the MAX phase [42]. Considering the double metal cations at M-site, Azina *et al.*, sputtered a $(Ti,Zr)_2AlC$ target on $Al_2O_3$(000*l*) substrate and found the formation of $(Ti,Zr)_2AlC$ solid solution MAX phase film as confirmed by HRTEM [43]. Interestingly, they also found that high substrate temperature (≥ 950 °C) is necessary to form single phase $(Ti,Zr)_2AlC$ thin film [44]. Mockute *et al.*, grew nanolaminated $(Cr,Mn)_2AlC$ thin films on $Al_2O_3$ (0001) by sputtering [45]. Meshkian *et al.*, synthesized high-quality $(Mo_{0.5}Mn_{0.5})_2GaC$ thin films on MgO (111) substrates, and performed the magnetic measurements showing a remanent magnetization ($M_r$) of ~0.35 $\mu_B$/M-atom at low temperature (hence ferromagnetism) [46]. This is the largest reported value among MAX compounds, till date. Later, $(Cr_{0.5}Mn_{0.5})_2GaC$ films were grown by Petruhins *et al.*, on MgO (111), 4H-SiC (0001), and $Al_2O_3$ (0001) with and without a NbN (111) seed layer, and they observed net magnetic moment of ~0.67 $\mu_B$/(Cr + Mn) atom at ~30 K and magnetic field of 5 Tesla [47]. Moreover, Salikhov *et al.*, measured the ferromagnetic resonance spectra of the same film, showing negligible magneto-crystalline anisotropy [48]. These excellent properties of 211 MAX phase thin films shows its enormous potential for wide-range of electrical, mechanical, thermal applications, even in high energy nuclear fusion reactors.

## 2. $M_3AX_2$ (*n*= 2)

312 MAX phase has become an important class of materials because of its unique combination of excellent ceramic and metallic properties**.** As seen, similar to the 211 phases, for the 312 phases, the metallic nature is governed mostly by the contribution of Ti-3*d* orbitals (**Fig. 3a**) [49]. Also, the most intriguing properties of 312 phase is the etching out the A-layers, forming new 2D material MXenes. Therefore, several efforts have been given to synthesis new 312 phases, not only in its bulk form but also in its high quality epitaxial thin film form, a viable need for the applications. At an early stage, Goto and Hirai grew monolithic $Ti_3SiC_2$ poly-crystalline plates by CVD [50]. Racault *et al.*, also attempted to grow $Ti_3SiC_2$ thin films by CVD, hoverer they obtained mixed phase film [51]. Later Jacques *et al.*, modified the growth method and used pulsed reactive CVD to grow films, however the results came out to be the same [52]. Pickering *et al.*, also tried to deposit $Ti_3SiC_2$ by CVD, forming a mixed phase compound [53]. First epitaxial single-crystalline $Ti_3SiC_2$ thin film growth was reported by Palmquist *et al.*, where they grew films on MgO (111) substrates by sputtering techniques, at ~900 °C [54]. They first grew a seed-layer of TiC (111), followed by the growth of $Ti_3SiC_2$ films, forming an epitaxy relation of $Ti_3SiC_2$ (0001)//TiC (111)//MgO (111), as confirmed by the cross-sectional TEM.

Emmerlich *et al.*, made an extensive structural characterizations about the formation of films. They also deposited few nm seed layer of $TiC_x$ (111) to grow epitaxial layered $Ti_3SiC_2$ thin films by sputtering, as confirmed by HRTEM (**Fig. 3b**) [55]. Interestingly, even without the pre-deposition, they also found that a thin layer of $TiC_x$ (111) was formed first on the MgO (111) substrate through the segregation of Si on top of the growth surface (**Fig. 3c**), relating to the complex growth kinetics and lattice relaxation mechanism, to minimize the high interfacial energy. Films with ~650 nm thickness shows very low electrical resistivity of ~25 μΩ-cm (on MgO (111)) and ~33 μΩ-cm (on $Al_2O_3$ (0001)). Films also show hardness of ~24 GPa and Young's modulus of ~343-370 GPa. Pécz *et al.*, found that formation of $Ti_3SiC_2$ phase via thermal evaporation of Al and Ti on *p*-type 6H-SiC (001) substrate and after annealing at ~900 °C in a nitrogen ambient, with films showing excellent Ohmic contact behavior [56]. Later, Tsukimoto *et al.*, did microstructural characterizations of this structure by TEM and found a hetero-epitaxial orientation relationship of (0001) $Ti_3SiC_2$ // (0001) SiC and (0-110) $Ti_3SiC_2$ // (0-110) SiC [57]. In addition, Fashandi *et al.*, also investigated the excellent Ohmic contact behavior of the same multilayer by growing $Ti_3SiC_2$ by using a single-step procedure by sputter deposition of Ti on 4H-SiC at ~960 °C [58]. Hopfeld *et al.*, measured the tribological properties of ~520 nm $Ti_3SiC_2$ thin films grown on Si (111) by magnetron sputtering and found the frictional coefficient value of ~0.15-0.25, making MAX useful as a lubricant material [35].

Hu *et al.*, grew ~200 nm $Ti_3SiC_2$ films by pulsed laser deposition on a polished silicon substrate at 100 °C and 300 °C and measured the mechanical properties, showing a friction coefficient of ~0.2 in humid air, and hardness of ~30-40 GPa [59]. Vishnyakov *et al.*, also deposited $Ti_3SiC_2$ at 650 °C by magnetron sputtering technique on native oxide covered Si substrate and did the Raman spectroscopy showing two active Raman mode of $E_{2g}$ (~622 $cm^{-1}$) and Ag (~657 $cm^{-1}$), arguing that this could be a fast identification of the formation of 312 MAX phases [60]. Recently, one major breakthrough was made in MAX was the intercalation of noble metals Au or Ir into the ~60 nm-thick $Ti_3SiC_2$ films grown on 4H-SiC by sputtering, forming the highly crystalline atomically ordered novel $Ti_3AuC_2$ and $Ti_3IrC_2$ MAX phases, as confirmed by the HRTEM (inset of **Fig. 3d**) [61]. Remarkably, these structures show excellent Ohmic contact with SiC which was even retained after ~1000 hrs of aging at ~600 °C (**Fig. 3d**), making MAX useful for numerous high temperature electronic device applications.

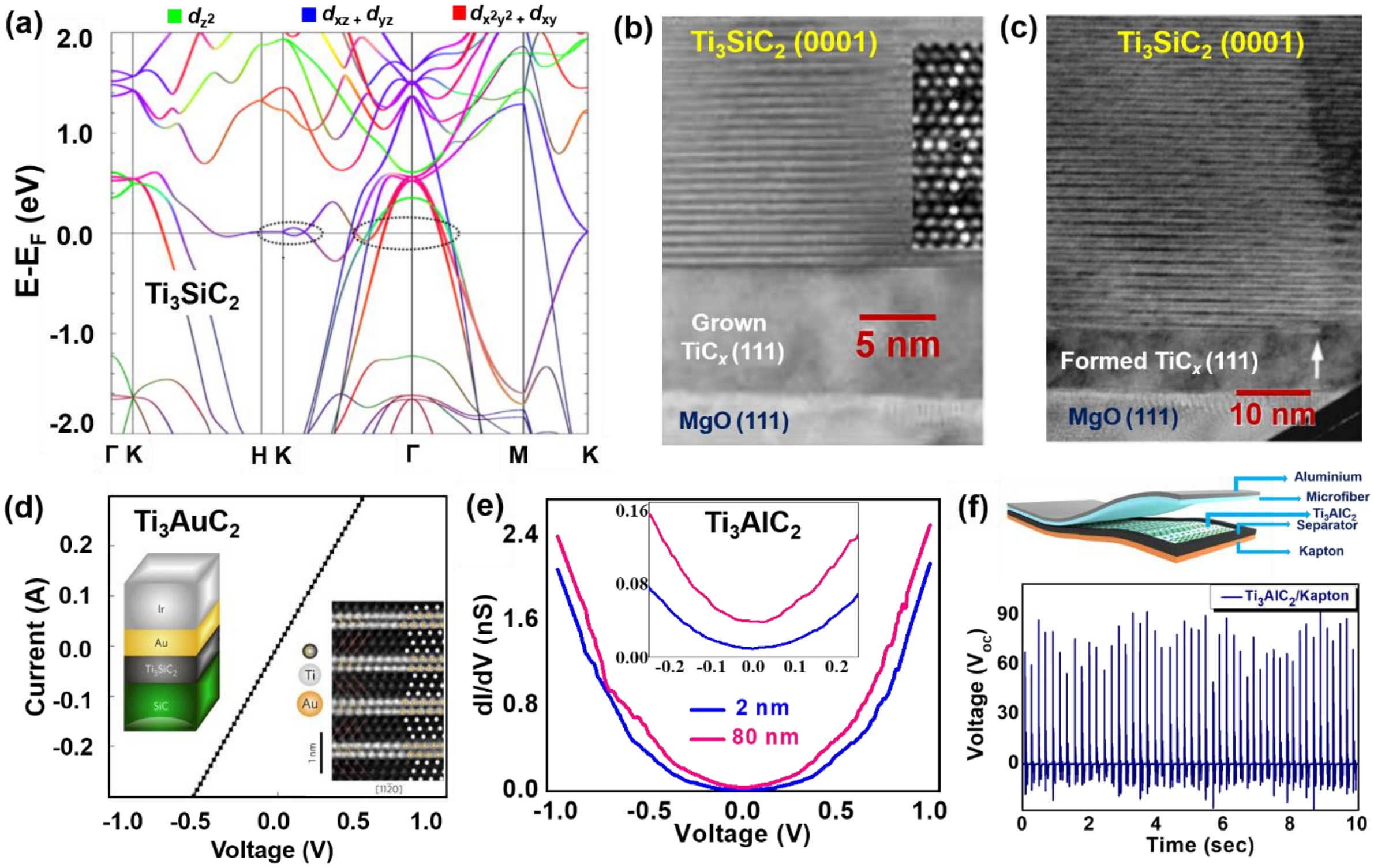

**Figure 3: Properties of $M_3AX_2$ (*n*=2) phases:** (a) Theoretical band structure of $Ti_3SiC_2$ shows metallic behavior with dominant contribution from the Ti-3*d* orbitals near the Fermi level. Atomic resolution TEM image of a layered $Ti_3SiC_2$ thin film with (b) an intentionally grown $TiC_x$ (111) seed layer, and (c) without grown but formed $TiC_x$ (111) seed layer during the deposition. (d) $Ti_3AuC_2$/SiC shows a perfect Ohmic contact behavior, with a linear *I-V* characteristics, which are retained even after annealing for ~1000 hrs. Inset shows the schematic of iridium and gold covered $Ti_3SiC_2$ film grown on SiC and the atomic resolution TEM image of the layered $Ti_3AuC_2$ thin film, forming ordered structure after noble-metal exchange of A layers while annealing at high temperature. (e) Room temperature scanning tunneling microscopy (STM) *I-V* curve of $Ti_3AlC_2$

films on $Al_2O_3$ (0001) shows the metallic characteristics. (f) Triboelectric nano-generator device fabrication by using a metallic $Ti_3AlC_2$ film on flexible Kapton substrate as an electrode showing the peak-to-peak open-circuit voltage ($V_{oc}$) of ~80V, by applying a human mechanical force of ~15 N. Adapted with permission from Pinek *et al.*, *Phys Rev B* 2020; **102**: 075111, copyright 2020 American Physical Society [**49**]; Emmerlich *et al.*, *J Appl Phys* 2004; **96**: 4817 [**55**]; Fashandi *et al.*, *Nat Mater* 2017; **16**: 814, copyright 2017 Macmillan Publishers Limited, part of Springer Nature [**61**]; Biswas *et al.*, *Phys Rev Appl* 2020; **13**: 044705, copyright 2020 American Physical Society [**65**].

Regarding other well-known 312 compounds, Wilhelmsson *et al.*, grew epitaxial $Ti_3AlC_2$ thin films by magnetron sputtering on $Al_2O_3$ (000*l*) substrates, with a ~20 nm $TiC_x$ (111) seed layer, showing resistivity, hardness and Young's modulus of ~51 μΩ-cm, ~20 GPa and ~260 GPa, respectively [**16**]. Magnuson *et al.*, deposited ~500 nm $Ti_3AlC_2$ films, ~ 200 nm $Ti_3SiC_2$ and $Ti_3GeC_2$ films by magnetron sputtering on $Al_2O_3$ (000*l*) substrate and investigated their electronic structures by soft x-ray emission spectroscopy [**62**]. Following the earlier reports, before the deposition of 312 phases, they also grew ~20 nm $TiC_{0.7}$ (111) the buffer layer to promote high quality growth. Interestingly, they found that the differences in the electronic structure are strongly related to the bonding nature between Ti-C and Ti-A layers. Later, they also investigated the substrates temperature growth of these films and found that films grown at room temperature possess almost ~2.6 at% of oxygen content [**63**], which deceases with the increase in growth temperature. The film also shows room temperature resistivity of ~120 μΩ-cm. Högberg *et al.*, also grew these three films via a similar technique on $Al_2O_3$ (000*l*) as well as MgO (111) substrates, which shows room temperature resistivity of ~25-30 μΩ-cm, slightly higher than the reported bulk phase (~22 μΩ-cm), as a consequence of the possible thickness reduction for film case [**25**]. Nearly ~600 nm $Ti_3AlC_2$ phase was also synthesized by Wilhelmsson *et al.*, via magnetron sputtering on $Al_2O_3$ (000l) and they found hardness of ~20 GPa and Young's modulus of ~240 GPa [**16**]. Halim *et al.*, grew ~15 to 60 nm $Ti_3AlC_2$ thin film on (000*l*) sapphire substrate by magnetron sputtering and showed that films are highly conducting with room temperature resistivity of ~35-45 μΩ-cm with transparency of less than 30% [**64**]. Remarkably, they etched out the Al layer from the film and formed the transparent (~90%) conducting $Ti_3C_2T_x$ MXene phase. Recently, we grew (103) oriented $Ti_3AlC_2$ thin film by pulsed laser deposition on $Al_2O_3$ (000*l*) substrates and investigated various functional properties, showing metallic behavior as confirmed by scanning tunneling microscopy (**Fig. 3e**), and its usefulness as Ohmic contact, and transparent ultrathin conductor [**65**]. Moreover, by using $Ti_3AlC_2$ film as an electrode material for tribological bio-sensor based devices fabrication (**Fig. 3f**), the device shows high open-circuit output voltage ($V_{oc}$) upon applied human force, which might be very useful for biomechanical touch sensors for energy harvesting. Very recently, Torres *et al.*, also synthesized $Ti_3AlC_2$ thin films by thermal treatment of a Ti-Al-C multilayer system via magnetron sputtering and found resistivity, hardness and Young's modulus of ~54 μΩ-cm, ~3.2 GPa, and ~120 GPa, respectively [**66**]. From magnetism point of view, Tao *et al.*, grew $(V,Mn)_3GaC_2$ thin films by sputtering. Interestingly, this is the first synthesized magnetic MAX thin film among the 312 phase. Film shows the hysteretic *M-H* loop between 50 K-300 K, indicating the possible presence of magnetic ordering well-above the room temperature [**67**] All these above observations point towards the remarkable application worthiness of 312 MAX phase thin films.

### 3. $M_4AX_3$ (*n*= 3) phase

Higher order (*n*= 3) 413 $M_4AX_3$ phase materials are also very interesting as they show high electrical conductivity, high damage tolerance, resistance to thermal shock, and high Young modulus. It possesses a very high out-of-plane lattice parameter of more than 20Å. Theoretical calculation shows the presence of a considerable density of states (DOS) at the Fermi level ($E_F$) (**Fig. 4a**) [**68**]. Bulk $Nb_4AlC_3$ shows metallic behavior with room temperature resistivity of ~75 μΩ-cm with (**Fig. 4b**), with Vickers hardness of ~2.6 GPa (**Fig. 4c**) [**69**]. The bulk $Ti_4AlN_3$ shows Young's moduli of ~310 GPa (**Fig. 4d**) [**70**]. $M_4AX_3$ phases are also useful as a coating material on spacecraft to avoid solar heating as well as an increase in radiative cooling [**68**] because of the increase in thermal emittance as compared to the TiN phase as reflectivity decreases dramatically in the infrared region and increases in the region of visible light, due to lower plasma frequency. An atomic-scale Z-contrast TEM microstructure image of a 413 phase of polycrystalline $(V_{0.5}Cr_{0.5})_4AlC_3$ shows the ABABACBCBC stacking (**Fig. 4e**) [**71**]. Therefore, it would be extremely interesting to synthesize the relatively complex higher-order $M_4AX_3$ phase materials, both in bulk as well as in thin-film form.

As found in the literature, few efforts have been given to grow this complex 413 phase. Högberg *et al.*, grew $Ti_4GeC_3$ thin films on $Al_2O_3$ (0001) substrates by dc magnetron sputtering at ~1000 °C which shows resistivity in the range of ~50-200 μΩ-cm [**24, 25**]. Emmerlich *et al.*, grew $Ti_4SiC_3$ thin films on $Al_2O_3$ (0001) at temperatures ranging from ~500 to 950 °C, by magnetron sputtering [**26**]. It was found that a thin layer of $TiC_x$ (111) is formed due to initial Si segregation, as confirmed by XRD. The film shows room temperature resistivity of ~51 μΩ-cm, with a nominal effect from the $TiC_x$ (111) seed layer at the film-substrate interface. Schramm *et al.*, obtained the $Ti_4AlN_3$ MAX phase by annealing the $(Ti_{1-x}Al_x)N_y$ thin films grown by reactive cathodic arc deposition [**72**]. They suggested that intercalation of Al and N along the basal plane accompanied by the simultaneous detwinning of the $Ti_2AlN$ crystal leads to the formation of a higher-order $Ti_4AlN_3$ phase. Recently, Li *et al.*, grew a new $Nb_4SiC_3$ MAX phase thin films by magnetron sputtering on Si (001) substrates and evaluate their properties by using several techniques [**73**]. Films were found to be highly conducting with room temperature resistivity of ~99 μΩ-cm. It also shows high hardness and elastic moduli values of ~15 GPa and ~200 GPa. Although the exploration of higher order $M_4AX_3$ phases is limited, probably because of the complexity in the synthesis of single-phase materials, however in the future, both theoretical insights about the growth and relevant experimental study will enrich the properties as well as possibly in tailoring the structure-property relationship in ever growing MAX phases.

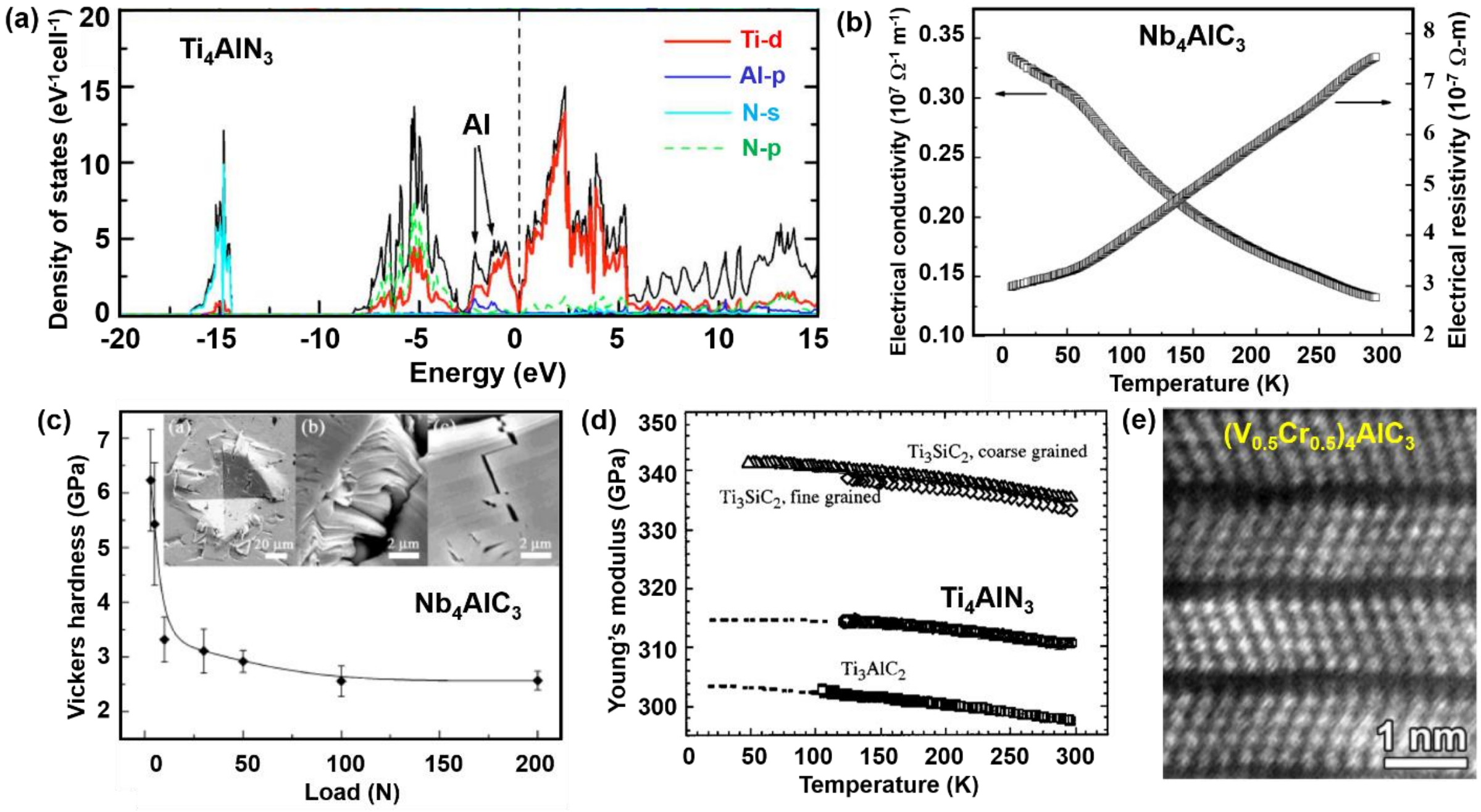

**Figure 4: Properties of $M_4AX_3$ (*n*=3) phases:** (a) Theoretical density of states (DOS) of $Ti_4AlN_3$. (b) Temperature dependent electrical resistivity of bulk $Nb_4AlC_3$ showing metallic behavior. (c) Vickers hardness as a function of indentation loads for bulk $Nb_4AlC_3$. Inset shows the SEM images at a different stage of the loads. (d) Temperature dependent Young's modulus of $Ti_4AlN_3$. (e) High-resolution Z-contrast TEM image of polycrystalline $(V_{0.5}Cr_{0.5})_4AlC_3$ along [1-210] direction. Due to the unavailability of high-quality epitaxial thin film growth and consequent functional properties, here we have only shown the excellent property potentials of bulk 413 MAX phase. Adapted with permission from Li *et al.*, *Appl Phys Lett* 2008; **92**: 221907, copyright 2008 American Institute of Physics [**68**]; Finkel *et al.*, *J Appl Phys* 2000; **87**: 1701, copyright 2000 American Institute of Physics [**70**]; Hu *et al.*, *J Am Ceram Soc* 2008; **91**: 2258, Zhou *et al.*, *J Am Ceram Soc* 2008; **91**: 1357, copyright 2008 The American Ceramic Society [**69, 71**].

## CONCLUSION AND OUTLOOK

In the last couple of decades several attempts have been made to grow high-quality epitaxial thin films of diverse (*n* = 1-3) MAX phases on various substrates by using different vapour phase deposition techniques (listed in **Table 2-4**). Films are mostly grown at high temperatures (~700-900 °C), an important parameter to grow high quality epitaxial single-phase MAX phase films. Remarkably, these films show very high electrical and thermal conductivity, high Vickers hardness and Young's modulus, excellent corrosion and oxidation resistance, low frictional force, high thermal and radiation damage tolerance that might be extremely useful for consequent application domains. Although substantial progress has been made, however considering the synthesis of a large number of bulk MAX phase compounds, thin film growth of MAX has a long way to go and there remain several difficulties and changes to solve in the future, and these are the following:

1. First of all, one needs to grow epitaxial single crystalline MAX phase thin films directly on structurally compatible substrates, without the pre-growth of a thin $TiC_x$ (111) seed layer, as it not only involves additional growth procedure but also contributes to the functional property measurements (however mere it is, having electrical resistivity of ~250 μΩ-cm), complicating in figuring out the actual contribution from the desired MAX film.
2. Low temperature growth without any surface oxidation is important as many times inter-diffusion between the elements from the substrate and of MAX film happens at a higher temperature, forming abrupt interfaces and mixed phases, hence deteriorating the film quality and consequent functional properties. In addition, growth of MAX films at different substrate temperatures would also be very interesting, considering the complex structures and consequent growth-temperature dependent changes in functional properties.

3. Considering the extensive lists of bulk MAX phases with highly anisotropic properties, it is always important and useful to grow new stable/metastable thin films, not only along out-of-plane (000*l*) orientations, but also along non-basal planes by epitaxy engineering, to widen the horizon of MAX with unforeseeable emergent properties.
4. MAX films are highly conducting as listed above, however tuning the band gap and making them semi-conducting-like and if possible, transparent too (via chemical doping/strain engineering/ultrathin films) would be important for the semiconductor industry as well as in photovoltaic technology.
5. Although some attempts have been given on finding magnetic MAX [**74**], however, growth of MAX thin films showing room-temperature ferro/anti-ferromagnetic ordering with high magnetic moment would be useful for spintronic applications.
6. The growth of ferroelectric MAX phase thin films would be useful for non-volatile memory device applications.
7. Growth of novel MAX phase films (e.g. high $T_c$ superconductors, topological insulators, quantum Hall effect etc.) for exotic quantum/topological applications.
8. Unfortunately, MAX lacks its application potential as a good catalyst for electrochemical hydrogen energy productions. Considering the cost-effectiveness, it would be very useful to engineer the film and making MAX as a high performance catalysts, which can eventually replace the expensive noble metal Pt or Ir catalyst, currently used for renewable energy productions.
9. Considering the surface oxidation problem, synthesizing new MAX phases, not only in thin film form but also as bulk materials containing the X-element other than C or N, would be of extreme importance. This looks a promising new direction as Boron (B) based material has been synthesized, named as MAB phase [**75**].
10. Finally, last but not the least, considering the advancement in thin film growth technology, in the future, integrating MAX with well-developed functional metal-oxide thin films, thus making a synergy between layered non-oxide and oxide heterostructures [**76**], would open up an entirely new avenue for emergent functionalities and technologically relevant next-generation applications.

**Table 2:** Growth and properties of $M_2AX$ ($n$=1) phase thin films

| Material | Method | Property | Reference |
|---|---|---|---|
| $Ti_2AlC$ | Magnetron sputtering | Anisotropic electrical resistivity | [15] |
| $Ti_2AlC$ | Magnetron sputtering | Low resistivity (~44 μΩ-cm) | [16] |
| $Ti_2AlC$ | Magnetron sputtering | Epitaxial growth | [17] |
| $Ti_2AlC$ | Pulsed cathodic arc | Epitaxial growth | [18] |
| $Ti_2AlC$ | Electron beam physical vapor deposition | Low friction coefficient | [19] |
| $Ti_2AlC$ | Magnetron sputtering | Epitaxial growth | [20] |
| $Ti_2AlC$ | Magnetron sputtering | Good oxidation resistance | [21] |
| $Ti_2AlC$ | Magnetron sputtering | Hexagonal laminal grains (~150-400 nm) | [22] |
| $Ti_2AlC$ | Magnetron sputtering | High temperature formed crystalline phase | [23] |
| $Ti_2GeC$ | Magnetron sputtering | Young's modulus (~300 GPa) | [24] |
| $Ti_2AlN$, $Ti_2AlC$ | Magnetron sputtering | Epitaxial growth | [25] |
| $Ti_2AC$ (A = Si, Ge, Sn) | Magnetron sputtering | Low resistivity (~20-50 μΩ-cm) | [26] |
| $Ti_2GeC$ | Magnetron sputtering | Low resistivity (~30 μΩ-cm), $p$-type carrier (~$3.5\times10^{27}$ $m^{-3}$) | [27] |
| $Ti_2AlN$ | Magnetron sputtering | Al diffusion at high temperature | [28] |
| $Ti_2AlN$ | Magnetron sputtering | High crystallinity and textured | [29] |
| $Ti_2AlN$ | Combined cathodic arc/sputter technique | Dense film with high-stability | [30] |
| $Nb_2AlC$ | Magnetron sputtering | Superconductivity at ~440 mK | [31] |
| $Cr_2GeC$ | Magnetron sputtering | Low resistivity (~53-66 μΩ-cm) with anisotropic electron-phonon coupling | [32] |
| $(Cr_{1-x}V_x)_2GeC$ | Magnetron sputtering | Metal-like electrical conductivity | [33] |
| $Cr_2AlC$ | Magnetron sputtering | Excellent mechanical properties | [34] |
| $Ti_2AlN$ and $Cr_2AlC$ | Magnetron sputtering | Friction coefficient (~0.15 to 0.50) and (~0.30 to 0.70) | [35] |
| $Cr_2AlC$ | PLD | Thickness dependent semiconductor-like and metallic-like behavior | [36] |
| $Cr_2AlC$ | Magnetron sputtering | Dual phase film with | [37] |
| | | radiation-tolerant | |
| $Cr_2AlC$ | Magnetron sputtering | Amorphous film and induced resistivity change | [38] |
| $V_2AlC$ | Magnetron sputtering | Structure analysis | [39] |
| $Mn_2GaC$ | Magnetron sputtering | Robust intra-layer ferromagnetic spin coupling | [40] |
| $Cr_2GaC$ | Magnetron sputtering | Thin film growth | [41] |
| $Ti_2AuN$ | Magnetron sputtering | Noble metals inter-diffusion | [42] |
| $(Ti,Zr)_2AlC$ | Magnetron sputtering | High temperature nucleation and growth of aluminides | [43] |
| $(Ti,Zr)_2AlC$ | Magnetron sputtering | High temperature for single phase growth | [44] |
| $(Cr,Mn)_2AlC$ | Magnetron sputtering | High-quality growth | [45] |
| $(Mo_{0.5}Mn_{0.5})_2GaC$ | Magnetron sputtering | Ferromagnetism at low T ($M_r$ ~0.35 $\mu_B$/M-atom) | [46] |
| $(Cr_{0.5}Mn_{0.5})_2GaC$ | Magnetron sputtering | Ferromagnetic behavior between 30-300 K ($M_s$ ~0.67 $\mu_B$/(Cr + Mn)) | [47] |
| $(Cr_{0.5}Mn_{0.5})_2GaC$ | Magnetron sputtering | Absence of magneto-crystalline anisotropy | [48] |

**Table 3:** Growth and properties of $M_3AX_2$ ($n$=2) phase thin films

| Material | Method | Property | Reference |
|---|---|---|---|
| $Ti_3AlC_2$ | Magnetron sputtering | Low resistivity (~51 μΩ-cm), high hardness (~20 GPa), Young Modulus (~260 GPa) | [16] |
| $Ti_3AlC_2$ | Magnetron sputtering | Epitaxial growth | [17] |
| $Ti_3GeC_2$ | Magnetron sputtering | Low resistivity (~50 μΩ-cm) | [24] |
| $Ti_3SiC_2$, $Ti_3GeC_2$, $Ti_3AlC_2$ | Magnetron sputtering | Low resistivity (~30-39 μΩ-cm), hardness (~15-25 GPa) | [25] |
| $Ti_3SiC_2$ | Magnetron sputtering | Friction coefficient (~0.15 -0.25) | [35] |
| $Ti_3SiC_2$ | CVD | Ploy-crystalline plates | [50] |
| $Ti_3SiC_2$ | CVD | Mixed phase | [51] |
| $Ti_3SiC_2$ | Reactive CVD | Mixed phase | [52] |
| $Ti_3SiC_2$ | CVD | Mixed phase | [53] |
| $Ti_3SiC_2$ | Magnetron sputtering | Epitaxial growth | [54] |
| $Ti_3SiC_2$ | Magnetron sputtering | Low resistivity (~25 μΩ-cm), hardness (~24 GPa), modulus (~343-370 GPa) | [55] |
| $Ti_3SiC_2$ | CVD | Ohmic contact | [56] |
| $Ti_3SiC_2$ | CVD | Atomic structure | [57] |
| $Ti_3SiC_2$ | Magnetron sputtering | Ohmic contact | [58] |
| $Ti_3SiC_2$ | PLD | Low friction coefficient (~0.2 in humid air), and high hardness (~30 and 40 GPa) | [59] |
| $Ti_3SiC_2$ | Magnetron sputtering | Structural and Raman spectroscopy | [60] |
| $Ti_3AuC_2$, and $Ti_3IrC_2$ | CVD | High temperature stable Ohmic contact | [61] |
| $Ti_3AC_2$, (A= Al, Si, Ge) | Magnetron sputtering | Weak covalent bonding and charge transfer | [62] |
| $Ti_3SiC_2$ | Magnetron sputtering | Low resistivity (~120 μΩ-cm) | [63] |
| $Ti_3AlC_2$ | Magnetron sputtering | Low resistivity (~37-45 μΩ-cm) | [64] |
| $Ti_3AlC_2$ | PLD | Bottom electrode, Schottky diode, Ohmic contact, transparent conductor, biomechanical touch sensor | [65] |
| $Ti_3AlC_2$ | Magnetron sputtering | Low resistivity (~54 μΩ-cm), high hardness (~3.2 GPa) | [66] |
| $(V,Mn)_3GaC_2$ | Magnetron sputtering | Ferromagnetic between 50 to 300 K | [67] |

**Table 4:** Growth and properties of $M_4AX_3$ ($n$=3) phase thin films

| **Material** | **Method** | **Property** | **Reference** |
|---|---|---|---|
| $Ti_4GeC_3$ | Magnetron sputtering | Low resistivity (~50 μΩ-cm) | [24] |
| $Ti_4SiC_3$ | Magnetron sputtering | Low resistivity (~51 μΩ cm) | [26] |
| $Ti_4AlN_3$ | Reactive cathodic arc deposition | Micro-structural analysis | [72] |
| $Nb_4SiC_3$ | Magnetron Sputtering | Low resistivity (~99 μΩ-cm), high hardness (~15 GPa), Young Modulus (~200 GPa) | [73] |

## ACKNOWLEDGMENTS

A.B. would like to acknowledge the hospitality at RICE University.

## AUTHORS' CONTRIBUTIONS

The manuscript was conceptualized by A.B. and written through the contributions of all authors. All authors have approved the final version of the manuscript.

## CONFLICT OF INTEREST STATEMENT

There are no conflicts of interest to declare.